\documentclass[journal,draftclsnofoot,onecolumn,12pt]{IEEEtran}

\usepackage[T1]{fontenc}

\usepackage{cite}
\usepackage[colorlinks,
linkcolor=red,
anchorcolor=blue,
citecolor=green
]{hyperref}

\ifCLASSINFOpdf
\else
\fi
\usepackage{amsmath}
\usepackage{diagbox}

\usepackage{ amssymb }
\usepackage{amsfonts}
\usepackage{caption}
\usepackage{graphicx}
\usepackage{lettrine}
\usepackage{epstopdf}
\usepackage{textcomp,booktabs}
\usepackage[usenames,dvipsnames]{color}
\usepackage{colortbl}
\definecolor{mygray}{gray}{.9}
\definecolor{mypink}{rgb}{.99,.91,.95}
\definecolor{mycyan}{cmyk}{.3,0,0,0}
\usepackage{pifont}
\usepackage{subfig}
\usepackage{multirow}
\usepackage{url}
\usepackage{color}
\usepackage{threeparttable}
\usepackage{setspace}
\usepackage{lineno}

\usepackage[dvipsnames,usenames]{color}

\usepackage{bm}

\usepackage{algorithm} 
\usepackage{algorithmic} 

\usepackage{dblfloatfix}

\usepackage[dvipsnames,usenames]{color}
\usepackage[usenames,dvipsnames]{color}

\definecolor{light-gray}{gray}{0.90}
\begin{document}
	\title{Deep  Source-Channel Coding for Sentence Semantic Transmission with HARQ }

	\author{Peiwen Jiang, Chao-Kai Wen, Shi Jin, and Geoffrey Ye Li
			\thanks{P. Jiang and S. Jin are with the National
				Mobile Communications Research Laboratory, Southeast University, Nanjing
				210096, China (e-mail: PeiwenJiang@seu.edu.cn; jinshi@seu.edu.cn).}
			\thanks{C.-K. Wen is with the Institute of Communications Engineering, National
				Sun Yat-sen University, Kaohsiung 80424, Taiwan (e-mail: chaokai.wen@mail.nsysu.edu.tw).}
			\thanks{G. Y. Li is with the Department of Electrical and Electronic Engineering,
				Imperial Colledge London, London, UK (e-mail: geoffrey.li@imperial.ac.uk).}}
	
	\maketitle
%
%
%
\begin{abstract}
		
Recently, semantic communication has been brought to the forefront  because of its great success in deep learning (DL), especially  Transformer. Even if semantic communication  has  been successfully applied in the sentence transmission to reduce  semantic errors,   existing architecture is usually fixed in the codeword length and is inefficient and inflexible for the varying sentence length. In this paper, we exploit hybrid automatic repeat request (HARQ) to  reduce semantic transmission error further. We first combine  semantic coding (SC) with  Reed Solomon (RS) channel coding and HARQ, called SC-RS-HARQ, which  exploits the superiority of the  SC and the reliability of the conventional methods successfully. Although the SC-RS-HARQ is easily applied in the existing  HARQ systems, we also  develop an end-to-end architecture, called SCHARQ, to pursue the performance further. Numerical results demonstrate that SCHARQ significantly reduces the required number of bits for sentence semantic transmission and sentence error rate. Finally, we attempt to replace  error detection from cyclic redundancy check to a similarity detection network called Sim32 to allow the receiver to reserve the wrong sentences with similar semantic information and to save transmission resources.

	\end{abstract}
	
	
	\newpage
	
	\setlength{\baselineskip}{22pt}
	\section{Introduction}

 	\IEEEPARstart{D}{eep} 
 learning (DL) has been utilized  in the physical layer of communications\cite{8054694,qin2018Deep,DBLP:journals/corr/abs-1809-06059}. DL-based joint design \cite{FCDNN,jointpilot,elbir2019online,ma2020sparse,guo2021canet,D2018Deep,ye2020deep,ye2021deep,9018199} has become a  potential directions to outperform the conventional communication structure.  In \cite{FCDNN},  a fully connected network is used to replace channel estimation (CE) and signal detection at the  receiver.  In \cite{jointpilot}, the  pilot  is jointly designed with CE to reduce  its overhead. In addition, the receiver  can also be jointly  optimized  with beamforming \cite{elbir2019online} and precoding \cite{ma2020sparse}. In \cite{guo2021canet}, the    entire network is combined with different modules,  including  pilot design, CE, and channel information state feedback.   Many modules in the conventional communication systems can  be jointly designed to improve their performance, such as DL-based joint encoder-decoder \cite{D2018Deep, ye2020deep,ye2021deep}. However, their  reliability still has  room for  improvement. In the conventional communication system,  reliable transmission is commonly guaranteed by
 hybrid	automatic repeat request (HARQ). Among different types of HARQ, incremental redundancy HARQ (IR-HARQ)  adjusts its code rate according to  the acknowledgment (ACK) feedback and is  a channel adaptive method. In  \cite{HARQ1}, the performance of IR-HARQ  achieved ergodic channel capacity for fading  channels. The existing DL-based  methods \cite{DLHARQ1,DLHARQ2} are proposed to enhance the HARQ but  are still designed independently. 
	
	Recently,  semantic communications \cite{strinati20216g,8476247,shi2021new} have been brought to the forefront because the  great success of DL makes many semantic tasks possible\cite{cho2014learning}.  Traditional communication systems concentrate on the bit or symbol level performance.  Semantic communication focuses on  transmitting the desired meaning, which is regarded as the second level communication in \cite{shannon1950mathematical}. Semantic communication usually goes beyond traditional Shannon’s paradigm because the separated source and channel coding approach is not always optimal in practice.  The semantic counterparts of Shannon’s source and channel coding theorems have been investigated in  \cite{bao2011towards}, where  semantic communication is content-related. Thus, shared and local knowledge can help the joint design of source and channel coding and  improve  transmission efficiency.  Specifically,  joint source and channel coding for semantic transmission has been  effective for image \cite{davis1996joint,bursalioglu2013joint,DJSCCC,yang2021deep,bourtsoulatze2019deep,8731945}, video \cite{zhai2005joint},  speech \cite{weng2021semantic}, and text \cite{farsad2018deep,xie2020deep} transmissions. 
	
	Some  architectures for DL-based semantic communications can  outperform those based on traditional communications for certain semantic metrics, especially when communication resources are limited and  noise is high. At the outset, fully connected and convolutional neural networks have been initially  exploited for some specific transmission tasks, such as image transmission \cite{bourtsoulatze2019deep,8731945}.   DL-based joint source and channel coding in \cite{bourtsoulatze2019deep} improves the peak signal-to-noise ratio while higher image classification accuracy is focused in \cite{8731945}. The shared knowledge of the transmission task is implicitly stored  in the trained weights of the neural networks and makes the DL-based semantic methods better than the  conventional  encoding methods, which exploit no knowledge of semantics.  A recurrent neural network (RNN) in \cite{farsad2018deep} is exploited for sentence transmission under the erasure channel, which  demonstrates its superiority under a high bit drop rate.   The semantic transceiver in \cite{xie2020deep}, which is called DeepSC, is based on Transformer \cite{vaswani2017attention} and is proved to be  better  than RNN in understanding the meaning of text sentences. DeepSC is extended to the  Internet of Things in \cite{9252948}, where more practical issues, such as channel impact, quantization, and network compression, are considered.  Furthermore,   the novel architecture based on federated edge intelligence in \cite{shi2020semantic} allows the user to offload semantic  tasks. However, the impact of  semantic coding   on  HARQ  still need further  study.
	
	In this paper, we  concentrate on  SC  based on Transformer for  semantic transmission of text sentences. However, IR-HARQ is introduced for further performance improvement, which is different from previous works. First, our semantic coding network for source coding, called SC,    is combined with the conventional  Reed-Solomon (RS) channel coding and HARQ. Then, an end-to-end autoencoder, called SCHARQ, is introduced to improve the transmission efficiency and reduce the sentence-error rate (SER) under a high bit-error-rate (BER). Finally, we replace the conventional error detection method  with a Transformer-based network, called Sim32,  to detect the meaning error in the estimated sentences.	The major contributions of this work are summarized as follows:
	
	1)  To improve the reliability of sentence semantic transmission, we combine the semantic source coding (SC) with conventional RS channel coding and HARQ, called parallel SC-RS-HARQ. The code length of SC is changeable according to the length of the sentences, thereby enabling SC to perform better than the methods with fixed code length when the required average number of bits is the same.   The proposed SC helps reduce SER under high BER but always has  minimal wrong words under low BER because DL has no mechanism to guarantee its performance when testing. The proposed parallel SC-RS-HARQ exploits the advantages of the semantic architecture and the conventional RS code and outperforms the competing semantic-based and conventional methods at word-error rate (WER), SER, and bit consumption.
	
	2) Although the proposed SC is easily  applied in the conventional HARQ systems because only source coding is replaced, we also try to reduce the SER and the required bits when transmitting a long sentence further in an end-to-end manner. A Transformer-based joint  source and channel coding, called SCHARQ, is proposed. SCHARQ  is more flexible than the existing RNN and Transformer-based sentence transmission methods because it can transmit incremental bits according to HARQ.   The code length of SCHARQ is controlled by the requirement of the receiver, which is more efficient than the competing methods when transmitting  sentences with different lengths. Meanwhile, SCHARQ  can   cope with high BER better than the separate design.
	
	3) To fully exploit the potential of the proposed semantic coding methods, we introduce a network called Sim32 to detect the meaning error in the received sentences. This error detection enables the received sentences to tolerate some error words as long as their meaning is unchanged. Sim32 can save  transmit resources because  more lossy sentences can be received without requiring retransmission.   However, the proposed error detection still makes  mistakes. For example, the replacement of nouns  may be ignored by  Sim32.

	The remainder of this paper is organized as follows. Section \uppercase\expandafter{\romannumeral2} introduces the system model, including conventional RS encoder,  IR-HARQ, and classic DL-based autoencoder architectures. The proposed networks are shown and discussed in Section \uppercase\expandafter{\romannumeral3}. In Section \uppercase\expandafter{\romannumeral4}, we demonstrate the superiority of the proposed networks in terms of SER and the required number of bits. Finally,  Section \uppercase\expandafter{\romannumeral5} concludes this paper.
	
\setlength{\baselineskip}{20pt}	
	\section{System Model and Related Works}
In this section, we first introduce a HARQ method for sentence transmission based on RS code. Next, we describe some existing DL-based  end-to-end transmission methods.  Finally, we  discuss the challenges of combining the deep semantic networks with the HARQ system. 
\subsection{HARQ System for Sentence Transmission}
To transmit a sentence, $\mathbf{s}$,  the conventional transmitter first converts it into bits by source and channel coding, thereby yielding
\begin{equation}
	\mathbf{b}=C_{\beta}(S_{\alpha}(\mathbf{s}))
\end{equation}
where $S_{\alpha}(\cdot)$ and $C_{\beta}(\cdot)$ denote the source encoder through $\alpha$ algorithm and channel encoder through $\beta$ algorithm, respectively.  $\hat{b}$ is denoted as the recovered bits at the receiver. Due to channel distortion, the recovered bits at the receiver may be different from those at the transmitter. 
The received sentence is decoded as 

\begin{equation}
\hat{\mathbf{s}}=S^{-1}_{\alpha}(C^{-1}_{\beta}(\hat{\mathbf{b}}))
\end{equation}
where $S^{-1}_{\alpha}(\cdot)$ and $C^{-1}_{\beta}(\cdot)$ represent the source and channel decoders with the corresponding algorithms, $\alpha$ and $\beta$, respectively.

Given the adaptive  correction capability of  IR-HARQ, it is  commonly used in communications, especially in wireless communication.  
Here, we consider an RS code, a maximum distance separable (MDS) code.  An $n$-symbol RS code with $k$ symbols of information can correct $n-k$ erasure symbols or $(n-k)/2$ error symbols. The corresponding  code rate can be calculated by $\frac{k}{n}$.  RS codes can be easily used for IR-HARQ because a punctured MDS code is still an MDS code \cite{MDS,RSMDS}. For example, $n$ total symbols are punctured to $n'$ $(<n)$. Correction capability becomes $n'-k$ erasure symbols or $(n'-k)/2$ error symbols.  Thus, if transmitting $n'$ symbols is not enough to correct the error symbols, we can transmit $n-n'$ incremental symbols to  increase its correction capability  from $(n'-k)/2$  to  $(n-k)/2$ error symbols.  
\begin{algorithm}[h]
	\caption{ HARQ transmission of sentences using RS channel coding.}
	
	\textbf{Input:} The transmitted sentence $\mathbf{s}$ \\ 
	\begin{enumerate} 

		\item Choose a low rate RS code. The full length of the codewords is $n_R$ and the number of information bits is $k$, yielding
		\begin{equation}
		\mathbf{b}_R\leftarrow C_{RS_{n_R}}(S_{\alpha}(\mathbf{s})).
		\end{equation}
		\item Puncture the encoder matrix and obtain different length of codewords, $\mathbf{b}_{1},\mathbf{b}_{2},\mathbf{b}_{3},\cdots$, where $k<n_1<n_2<n_3<\cdots<n_R$.
		\item Transmit $\mathbf{b}_{1}$ and 32 CRC parity bits.
		\item $\hat{\mathbf{s}}_1\leftarrow S^{-1}_{\alpha}{C^{-1}_{RS_{n_1}}(\hat{\mathbf{b}}_{1}})$.\\
		\item   \textbf{if} CRC detects error, feedback NACK; \textbf{else}  $\hat{\mathbf{s}}=\hat{\mathbf{s}}_1$ and feedback ACK.\\

	\end{enumerate}
\textbf{for} $i =2; i\leqslant R ; i++$ \textbf{do}\\
\hspace*{0.3cm} \textbf{if} NACK\\
	\hspace*{0.6cm}	1) Transmit the incremental symbols, $\mathbf{b}_{i}-\mathbf{b}_{{i-1}}$.\\ 
 \hspace*{0.6cm}	2) $\hat{\mathbf{s}}_i\leftarrow S^{-1}_{\alpha}{C^{-1}_{RS_{n_i}}(\hat{\mathbf{b}}_{i}})$.\\ 
 \hspace*{0.6cm}   3) \textbf{if} CRC detects error, feedback NACK; \textbf{else}  $\hat{\mathbf{s}}=\hat{\mathbf{s}}_i$ and feedback ACK.\\
 \hspace*{0.3cm} \textbf{end if} \\ 
 \textbf{end for} \\
\textbf{if}  ACK	\\	
\hspace*{0.3cm}	\textbf{Output:} $\hat{\mathbf{s}}$. \\
\textbf{else}\\
\hspace*{0.3cm}	\textbf{Output:} Sentence error. \\	
	\label{RSHARQ}
\end{algorithm}

The HARQ transmission process for sentences is described in Algorithm \ref{RSHARQ}. If the estimated sentences at the receiver are detected to be correct by cyclic redundancy check (CRC) error detection, the  acknowledged (ACK) signal will be sent to the transmitter, and the reserved incremental symbols are not  required. In contrast, the sentence is unsuccessfully received when all  incremental symbols are transmitted, but the full-length codewords still cannot be decoded correctly.
\subsection{DL-based Autoencoder}
The joint design has a  great potential to improve  transmission efficiency. For the DL-based autoencoder,
the conventional encoder and decoder,  $C_{\beta}(S_{\alpha}(\cdot))$ and $S^{-1}_{\alpha}(C^{-1}_{\beta}(\cdot))$, are replaced by the DL-based encoder and decoder. 
To train the encoder-decoder architecture in an end-to-end manner, they are connected by a channel layer, which  usually consists of a dropout layer and an AWGN layer if the quantization is not considered.  The channel layer can  be a bit-erasure or bit-error layer for the quantized encoder and decoder. Also, the channel layer can learn the fading channel through the generative adversarial network\cite{ye2020deep}.

The DL-based autoencoders  perform better than the conventional methods, especially under  wired environments with nonlinear interference and limited transmission resources. The joint source-channel coding \cite{farsad2018deep} initiates the words with Glove
pre-trained embeddings \cite{pennington2014glove} and uses RNN to learn the semantic information. In \cite{xie2020deep},  the attention mechanism is used for the semantic coder, which is called Transformer in \cite{vaswani2017attention}. The first step of the semantic encoder for a sentence is to obtain the word embedding.  The sentence with $L_{\mathbf{s}}$ words can be expressed by a positive integer vector, $\mathbf{s}=[w_1,w_2,\cdots,w_{L_{\mathbf{s}}}]$. As the input of a DL-based encoder, all the sentences are zero-padded to a length of $L$. The word embedding process needs  a lookup table $\mathbf{\Psi}$ and $L_\mathbf{\Psi}$ 
words in the dictionary. Denote $M$ as the length of the word vectors after the word embedding process. Thus, $\mathbf{\Psi}$ is an  $L_\mathbf{\Psi}\times M$ real matrix whose parameters are trainable. The word embedding process is denoted as $f_{\rm embed}(\cdot;\cdot)$, and  a sentence after wording embedding can be written as
\begin{equation}
\mathbf{V}=f_{\rm embed}(\mathbf{s};\mathbf{\Psi})=\left[ \begin{matrix}
	\mathbf{\Psi}[w_{1}]	\\
	\vdots	\\
	\mathbf{\Psi}[w_L]\\
\end{matrix} \right]+\mathbf{PE}
\end{equation}
where $\mathbf{V}\in \mathbb{R}^{L\times M}$, $\mathbf{s}$ is the input sentence,   $\mathbf{\Psi}[w_{1}]$ is the vector of the $w_{1}$-th row in the trainable  $\mathbf{\Psi}$, and the  additive  matrix $\mathbf{PE}$ is a constant   matrix for position encoding  defined in \cite{vaswani2017attention}. These word vectors in  $\mathbf{\Psi}$ contain the meaning of the words because the distance of any two similar-word vectors is usually shorter than that of dissimilar-word vectors.  Pre-trained lookup tables are available for extracting semantic information,  such as Word2Vec\cite{mikolov2013efficient} and Glove\cite{pennington2014glove}.    The detailed architecture of the Transformer can  be referred to  \cite{vaswani2017attention}. For convenience, we denote the Transformer-based encoder  and decoder  as $T_{\rm en}(\cdot)$ and $T_{\rm de}(\cdot)$, respectively. Here, the  trainable parameters in these processes are not shown explicitly, and $f_{\rm embed}(\mathbf{s};\mathbf{\Psi})$ is also simplified to $f_{\rm embed}(\mathbf{s})$.  An FC layer converts the output of Transformers in the decoder  from $\mathbb{R}^{L\times M}$ to $\mathbb{R}^{L \times L_\mathbf{\Psi}}$ via the  SoftMax activation.  Thus, the decoded sentence  $\hat{\mathbf{s}}$ is obtained according to the index of the maximum value at each row. The process  is denoted as $f_{\rm argmax}(\cdot)$. For machine translation, the translated sentence $\hat{\mathbf{s}}$ can be written as 
\begin{equation}
\hat{\mathbf{s}}=	f_{\rm argmax}(T_{\rm de}(T_{\rm en}(f_{\rm embed}(\mathbf{s})))).
\end{equation}
For semantic communication, several FC layers\cite{xie2020deep} are used to compress $T_{\rm en}(f_{\rm embed}(\mathbf{s}))$ into transmit symbols, and the received symbols are also decompressed by FC layers as
\begin{equation}
	\hat{\mathbf{s}}=	f_{\rm argmax}(T_{\rm de}(f_{\rm de}(h(f_{\rm en}(T_{\rm en}(f_{\rm embed}(\mathbf{s}))))))), \label{eqsc}
\end{equation}
where $h(\cdot)$ is the channel layer, and $f_{\rm en}(\cdot)$ and $f_{\rm de}(\cdot)$ are  the processes of FC layers. Then, this architecture is trained to cope with the effect of channels.
  
The semantic networks achieve better performance than  conventional coders, especially when the new semantic metrics, such as BLEU\cite{papineni2002bleu}, are applied. However,  the benefits of the semantic methods for the throughout of a sentence transmission system are needed to be investigated further.

\subsection{Challenges on   Semantic Coders}
The most obvious issue for the semantic coders is the fixed network architecture for  different sentence lengths, which decreases  coding efficiency. The fixed bit  transmission in \cite{farsad2018deep} performs better for  short sentences than   long ones. The fixed symbol  transmission in  \cite{xie2020deep}  also faces a similar issue if we add the quantization module.

In addition, they  lack  combination with  HARQ, which is  vitally important for  successful transmission. The superiority of the semantic networks are exhibited  not only in lossless compression but also in lossy transmission. The  sentences with wrong words estimated by the semantic networks  also contain useful semantic information under extreme hostile environments. To adapt to this content-related encoder-decoder, the HARQ method should  also be refined.

\section{HARQ based on Semantic Coder}
In this section, we propose different semantic architectures, that are combined with IR-HARQ in different extents. In the beginning, the design of the semantic network is independent of the  HARQ framework. Then, all the source-channel coding and the incremental encoded bits are  generated by the neural networks. Finally, we  replace the CRC with the similarity detection to reserve the incorrect but similar  sentences at the receiver.
\subsection{Semantic Source Coding for Transmitter with RS Channel Coding}
Joint semantic source and channel coding have been studied in \cite{farsad2018deep,xie2020deep}. These existing methods have shown their superiority under low SNR and the limited number of bits for each sentence. However, the fixed architectures are not flexible and efficient because sentences  usually have different lengths. Meanwhile, these joint designs make the combination  channel coding and HARQ difficult because DL-based coding is inexplicable. 

\begin{figure}[!h]
	\centering

		\subfloat[ ]{
		\includegraphics[width=6in]{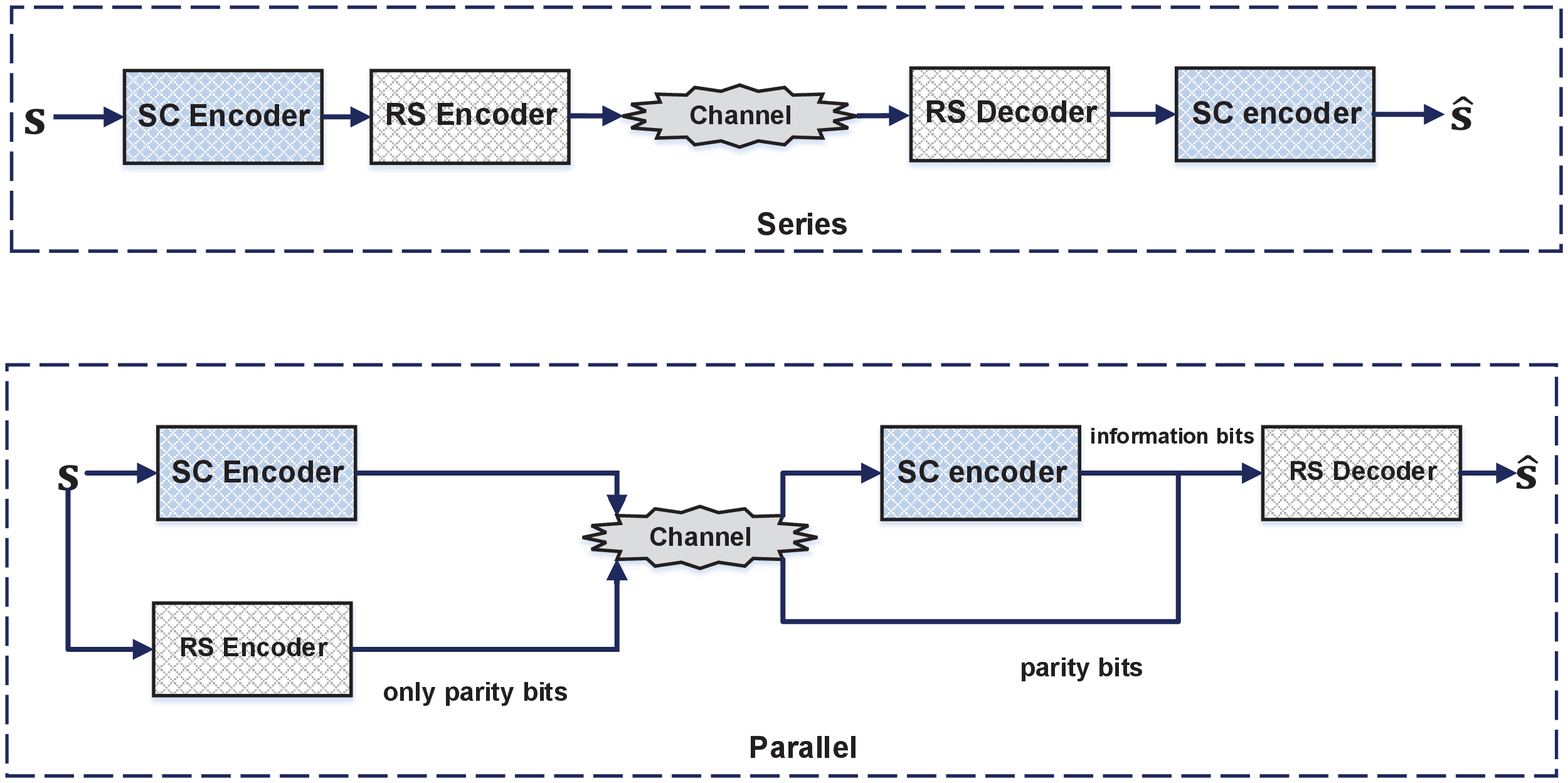}}\\
	\subfloat[ ]{
		\includegraphics[width=6in]{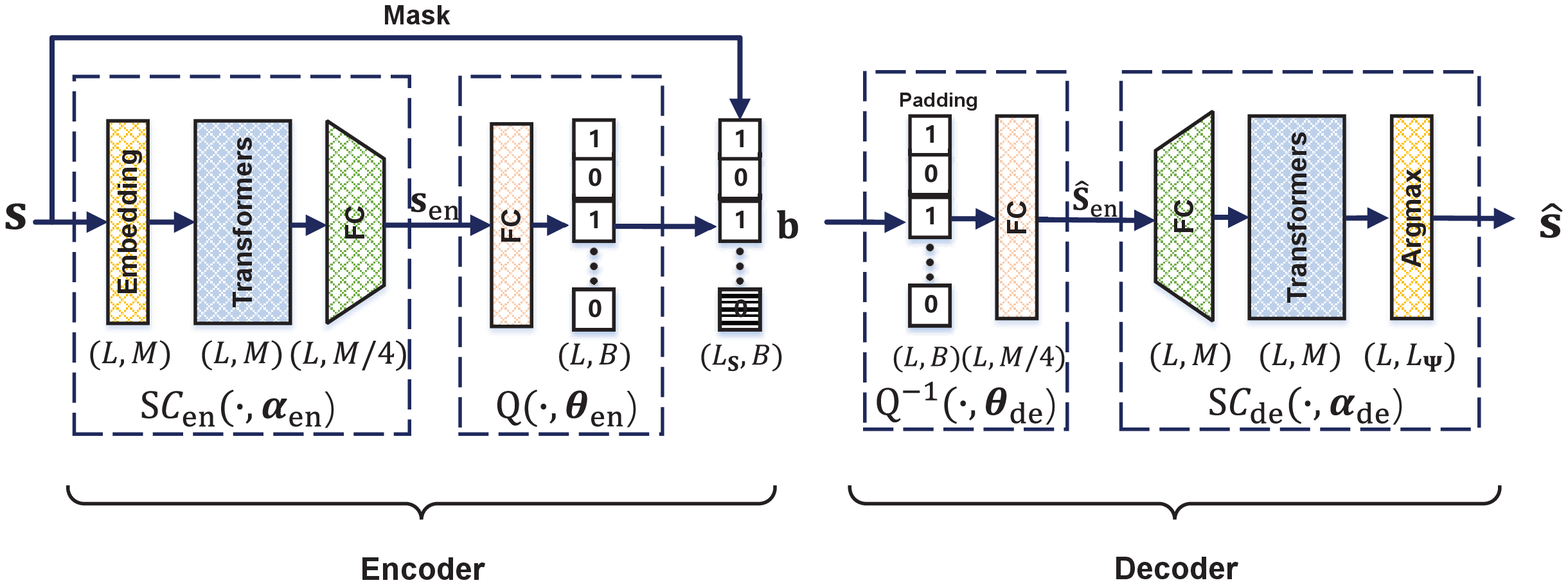}}
	
	\caption{(a) Two combination methods of SC- and RS- based IR-HARQ. (b) Architecture of the Transformer-based  SC. The encoded bits are masked according to the input sentence length. }
	\label{SC_coder}
\end{figure}
As shown in Fig. \ref{SC_coder}(a), we first  designed SC separately without the combination of  conventional RS coding and IR-HARQ, and  IR-HARQ (not shown in the figure) is directly based on the RS coding. The two proposed  methods are called series and parallel SC-RS-HARQ. These methods exploit  conventional RS coding in different stages. For series SC-RS-HARQ, the RS coding is used to protect the coded bits of the SC. That is, the received sentences are directly decoded by SC.  Thus, this method may still commit mistakes under no transmission errors  because the performance of the DL-based SC is not guaranteed when testing. For parallel SC-RS-HARQ, the sentence is encoded by SC and RS, and the parity bits of the RS coding are transmitted directly. Therefore, the RS coding is used to protect the original sentence and the  correct  sentence  decoded by the SC. Thus, the received sentences are directly decoded by  RS coding, and their performance is guaranteed under low BER.  Overall, these methods can achieve better performance under high BER due to the introduction of SC.

 In addition, we develop the SC to encode the sentence into different lengths of bits according to the length of this sentence (Fig. \ref{SC_coder}(b)), which is beneficial for code efficiency.   The encoder and decoder are based on Transformer \cite{vaswani2017attention}. The sentence is mapped to a vector of positive integers $\mathbf{s}=[w_1,w_2,\cdots,w_L]$ as an input. The SC encoder embeds $\mathbf{s}$ as  (4), and then the embedded word vectors are processed by six Transformers with $M$ units. The output of the Transformers is compressed by an FC layer, and the entire process is denoted as  $SC_{\rm en}(\cdot)$ with  output 
\begin{equation}
	\mathbf{s}_{\rm en}=SC_{\rm en}(\mathbf{s};{\bm \alpha_{\rm en}}),
\end{equation}
where $\mathbf{s}_{\rm en}\in \mathbb{R}^{L \times M/4}$ and ${\bm \alpha_{\rm en}}$ represent all the trainable parameters. The one-bit quantization module first converts $\mathbf{s}_{\rm en}$ into $\mathbb{R}^{L \times B}$ with an FC layer, where $B$ is the number of bits for each word.   Different from the existing methods, the quantization module also masks part of the bits in accordance to sentence length $L_{\mathbf{s}}$. We denote the quantization process as $Q(\cdot)$, and its output can be expressed as 
\begin{equation}
	\mathbf{b}=Q(\mathbf{s}_{\rm en};{\bm \theta_{\rm en}}),
\end{equation}
where $\mathbf{b}$ is an $L_\mathbf{s} \times B$ bit vector if the sentence length is $L_\mathbf{s}$. The other $(L-L_\mathbf{s}) \times B$ bits are not transmitted  to save  transmission resources.

The dequantization module pads input bits to an $L \times B$ binary matrix with zeros. Then, an FC layer is used for reshaping, and its output is $\mathbf{\hat{s}}_{\rm en}\in \mathbb{R}^{L \times M/4}$, thereby yielding 
\begin{equation}
	 \mathbf{\hat{s}}_{\rm en}=Q^{-1}({\mathbf{b}};{\bm \theta_{\rm de}}).
\end{equation}
Then, $\mathbf{\hat{s}}_{\rm en}$ is  decompressed into $\mathbb{R}^{L \times M}$ with an FC layer and then goes through six Transformers with $M$ units. At last, the argmax layer uses an FC layer with SoftMax activation and outputs $L \times L_\Psi$ vectors of probabilities in the dictionary. The estimated sentence is composed of the max possible words.  Similar to the encoder process, the estimated sentence  can be written as 
\begin{equation}
	\hat{\mathbf{s}}=SC_{\rm de} (\mathbf{\hat{s}}_{\rm en};{\bm \alpha_{\rm de}}).
\end{equation}

 The number of bits for each word, $B$, is difficult to choose to balance the coding efficiency and bit consumption; thus, an end-to-end training process costs too much time.  The entire training process is also divided into three steps.

1) The parameters in  $SC_{\rm en}(\cdot)$ and  $SC_{\rm de}(\cdot)$ are trained without quantization layers $Q(\cdot)$ and $Q^{-1}(\cdot)$, i.e., $\mathbf{\hat{s}}_{\rm en}=	\mathbf{s}_{\rm en} $. This training process can be expressed as 

\begin{equation}
	(\hat{\bm{\alpha}}_{\rm en},\hat{\bm{\alpha}}_{\rm de})=\mathop{\arg\min}\limits_{{\bm{\alpha}}_{\rm en},{\bm{\alpha}}_{\rm de}}L_{\rm CE}\left(\mathbf{s}, SC_{\rm de}(SC_{\rm en}(\mathbf{{s}};{\bm{\alpha}}_{\rm en});{\bm{\alpha}}_{\rm de})\right),
\end{equation}
where $L_{\rm CE}(\cdot)$ denotes the cross-entropy (CE) loss function. 

2) Try  $B$ repeatedly until  transmit bits are enough to carry the information in $\mathbf{s}_{\rm en}=SC_{\rm en}(\mathbf{{s}};\hat{\bm{\alpha}}_{\rm en})$. The loss function is the mean-squared error (MSE), and the training process can be expressed as

\begin{equation}
	(\hat{\bm{\theta}}_{\rm en},\hat{\bm{\theta}}_{\rm de})=\mathop{\arg\min}\limits_{{\bm{\theta}}_{\rm en},{\bm{\theta}}_{\rm de}}L_{\rm MSE}\left(\mathbf{s}_{\rm en}, Q^{-1}(Q(\mathbf{s}_{\rm en};\bm{\theta}_{\rm en});\bm{\theta}_{\rm de})  \right).
\end{equation}
3) Finetune all  trainable parameters in an end-to-end manner,
\begin{equation}
	(\hat{\bm{\alpha}}_{\rm en},\hat{\bm{\alpha}}_{\rm de},\hat{\bm{\theta}}_{\rm en},\hat{\bm{\theta}}_{\rm de})=\mathop{\arg\min}\limits_{{\bm{\alpha}}_{\rm en},{\bm{\alpha}}_{\rm de},{\bm{\theta}}_{\rm en},{\bm{\theta}}_{\rm de}}L_{\rm CE}\left(\mathbf{s}, \hat{\mathbf{s}}\right).
\end{equation}

The  two  methods design SC and conventional IR-HARQ modules separately so that  they can be   applied easily.  The advantages of SC are exploited under high BER.  Meanwhile, the conventional method shows its superiority under the lossless transmission, and this combination can cover the shortage of the AI method. 
However,  joint  optimization is also a potential strategy for reducing transmission resources further,   and thus joint source-channel coding and HARQ architecture will be studied further.

\subsection{Semantic-based End-to-End HARQ}
 The long sentences can also be dealt through transmitting incremental bits because of the adjustable length of IR-HARQ. Thus, we propose an end-to-end semantic framework similar to the IR-HARQ framework, and it is called SCHARQ, which transmits the incremental bits until the receiver estimates  sentence successfully  or reach the maximum number of retransmissions. These incremental bits can  not only improve correction capability but can also carry extra information for complex sentences. Overall, this framework is flexible under varying channels and different lengths of sentences.

\begin{figure}[!h]
	\centering
	
	\includegraphics[width=6.5in]{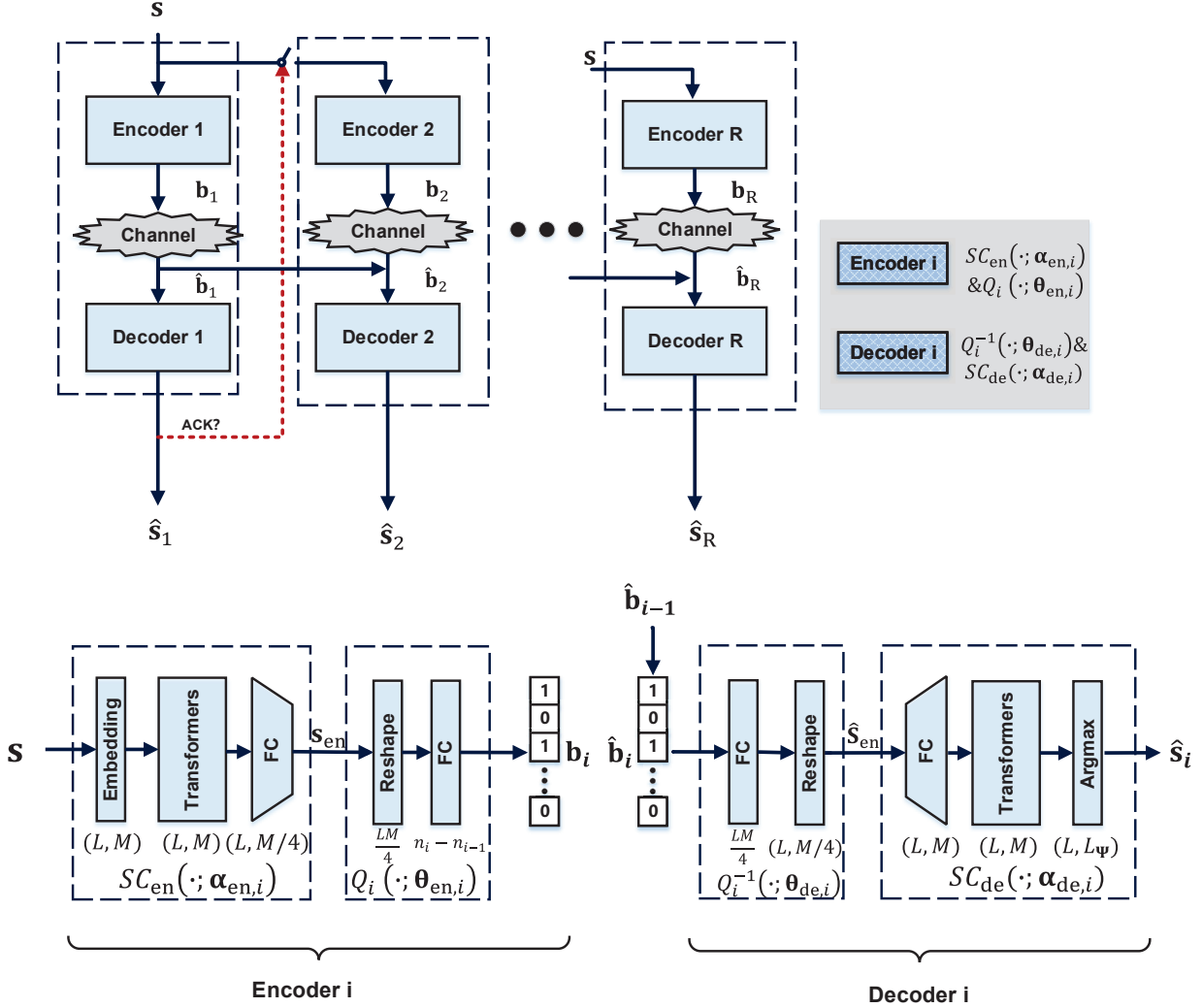}
	
	\caption{Structure of  SCHARQ, which can transmit $R$ times in maximum.   }
	\label{SCHARQar}
\end{figure}

There are $R$ SC-based encoders and decoders  for $R$ transmissions are shown in Fig. \ref{SCHARQar}. Here, the encoder and decoder architectures are different from those in Fig. \ref{SC_coder}. The quantization process $Q_i(\cdot;\bm \theta_{{\rm en},i})$ in SCHARQ does not need to  mask part of  bits, but it  converts the output of $SC_{\rm en}(\cdot;\bm \alpha_{{\rm en},i})$ to $\mathbb{R}^{1 \times (n_i-n_{i-1})}$ with a dense layer and then uses one-bit quantization.   An $n_1$-bit vector $\mathbf{b}_1$ is transmitted first, and the following transmissions use the similar architecture.  The $i$-th transmission can be expressed as 
\begin{equation}
\mathbf{b}_{i}=Q_i(SC_{\rm en} ({\mathbf{s}};\bm \alpha_{{\rm en},i});\bm \theta_{{\rm en},i}),
\end{equation}
and the previous transmitted bits are connected with the incremental bits $\hat{\mathbf{b}}_{i}$ and decoded together, yielding

\begin{equation}
	\hat{\mathbf{s}}_i=SC_{{\rm de}} (Q_i^{-1}([\hat{\mathbf{b}}_{1},\cdots,\hat{\mathbf{b}}_{i}];\bm \theta_{{\rm de},i});\bm \alpha_{{\rm de},i}), 
\end{equation}
where $Q_i^{-1}(\cdot)$ reshapes these bits with a dense layer  and $SC_{{\rm de}}(\cdot)$ has the same architecture as in Fig. \ref{SC_coder}.

The training process uses $\hat{\bm{\alpha}}_{\rm en}$ and $\hat{\bm{\alpha}}_{\rm de}$ trained in Section III A as the initiation of  ${\bm{\alpha}}_{{\rm en},i}$ and ${\bm{\alpha}}_{{\rm de},i}$, respectively.  For the first transmission, the training process is similar to  Steps 2 and 3 in Section II A, but the channel condition should be considered here, and 5\% bits are in error. For the $i$-th transmission ($i>1$), the trainable parameters of the previous transmissions are fixed and the training process can be expressed as  
\begin{equation}
	(\hat{\bm{\alpha}}_{{\rm en},i},\hat{\bm{\alpha}}_{{\rm de},i},\hat{\bm{\theta}}_{{\rm en},i},\hat{\bm{\theta}}_{{\rm de},i})=\mathop{\arg\min}\limits_{{\bm{\alpha}}_{{\rm en},i},{\bm{\alpha}}_{{\rm de},i},{\bm{\theta}}_{{\rm en},i},{\bm{\theta}}_{{\rm de},i}}L_{\rm CE}\left(\mathbf{s}, \hat{\mathbf{s}}_i\right), i >1.
\end{equation}
The application of this framework is naturally the same as the conventional IR-HARQ shown in Alg. 1. After the $i$-th transmission, we check $\hat{\mathbf{s}}_i$ with CRC and then transmit $\mathbf{b}_{i+1}$ if this transmission cannot obtain the correct sentence. If the $i$-th transmission passes the CRC error detection, then $\hat{\mathbf{s}}=\hat{\mathbf{s}}_i$, and the subsequent transmission is not required.

\subsection{Similarity Detection}
The above joint design with HARQ makes the network more flexible under varying channels and sentences, but  the system still aims at minimizing word error  rather than  the semantic error in transmission. To fully  exploit the  potential of semantic architecture, we attempt to change the CRC  to a similarity detection in this section. 

Conventional transmission systems usually rely on the CRC error detection to  repeat requests automatically to receive the correct sentences and feedback ACK. However, the traditional CRC error detection will regard a sentence with an error if it contains  error words but  with the same or similar meaning as the original one.  Similar sentences are also useful for semantic transmission, especially in hostile environments.

 Although some methods for the similarity measurement of sentences, such as Levenshtein distance and BLEU, have been proposed. They only calculate  the change of words between two sentences and have no insight into the meaning of different words. Recently, BERT\cite{devlin2018bert}, a pre-trained model under billions of words and sentences, achieved great success in extracting semantic information. This architecture has been  applied in measuring the similarity of sentences, such as \cite{xie2020deep}.  $BERT(\mathbf{s})$  converts the input sentences, $\mathbf{s}$, into  real vectors,  and the cosine similarity is used to measure the similarity in their semantic information, which is defined as

\begin{equation}
	Sim(\mathbf{s},\hat{\mathbf{s}})=\frac{BERT(\mathbf{s})BERT(\hat{\mathbf{s}})^T}{|BERT(\mathbf{s})| |BERT(\hat{\mathbf{s}})|}. \label{simcal}
			\end{equation}

\begin{figure}[!h]
	\centering
	
	\includegraphics[width=6.5in]{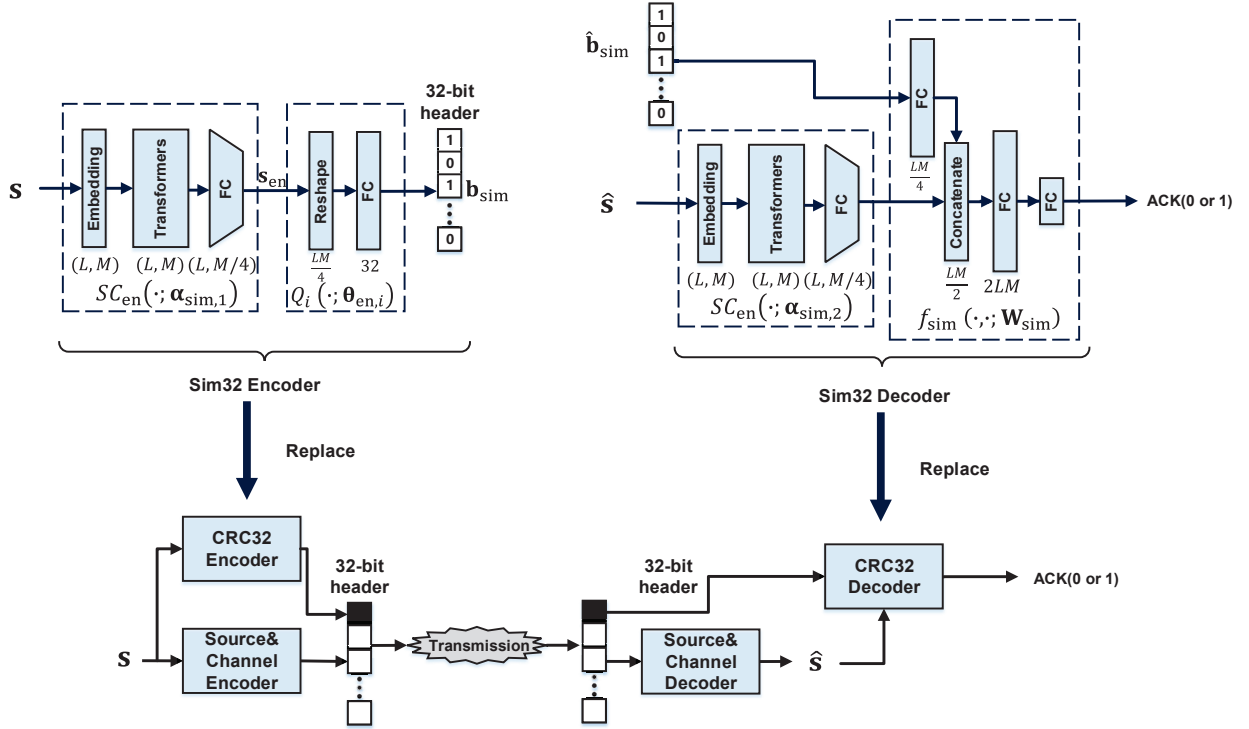}
	
	\caption{Structure of similarity detection method. The estimated sentence $\hat{\mathbf{s}}$  is  received by the proposed methods, such as SC-RS-HARQ and SCHARQ.  }
	\label{SimNet}
\end{figure}

However, the true sentence $\mathbf{s}$ in  HARQ systems is unavailable at the receiver. Thus, a new method that is similar to CRC is proposed, and it is called Sim32. In Fig. \ref{SimNet}, 32 bits are transmitted to the receiver for  similarity detection.  At the transmitter, the 32 bits is encoded as
\begin{equation}
	\mathbf{b}_{\rm sim}=Q_{\rm sim}(SC_{\rm en}(\mathbf{s}; \bm \alpha_{\rm sim, 1}); \bm \theta_{\rm sim,1}),
\end{equation}
  where $Q_{\rm sim}(\cdot;\bm \theta_{\rm sim,1})$ converts the output of $SC_{\rm en}(\cdot; \bm \alpha_{\rm sim, 1})$ into 32 bits, and $\bm \alpha_{\rm sim, 1}$ and $\bm \theta_{\rm sim,1}$ are the trainable parameters in these processes. At the receiver,  similarity detection is based on two inputs, namely, $\hat{\mathbf{b}}_{\rm sim}$ and $\hat{\mathbf{s}}$, which  can be expressed as 
 
\begin{equation}
	Sim32(\hat{\mathbf{s}},\hat{\mathbf{b}}_{\rm sim})=f_{\rm sim}(\hat{\mathbf{b}}_{\rm sim},SC_{\rm en}(\hat{\mathbf{s}}; \bm \alpha_{{\rm sim},2}); \mathbf{W}_{\rm sim}),
\end{equation}
where $ f_{\rm sim}$ has two FC layers and the trainable parameters, $\bm \alpha_{{\rm sim},2}$ and $\mathbf{W}_{\rm sim}$ are introduced. Its hidden layer  has  four times the size of the input and ReLU activation function, and it only outputs one value using the sigmoid activation function. 

The training process of this architecture will collect the estimated sentences from the aforementioned frameworks under different channel conditions and  retransmission stages. The label is based on  (19), and it satisfies

\begin{equation}
	label(\mathbf{s},\hat{\mathbf{s}})=\left\{
\begin{aligned}
1	, & & Sim(\mathbf{s},\hat{\mathbf{s}})>0.98, \\
0	, & & Sim(\mathbf{s},\hat{\mathbf{s}})\leq 0.98, \\
\end{aligned}
\right. 
\end{equation}
where  $Sim(\mathbf{s},\hat{\mathbf{s}})>0.98$ indicates that the estimated sentences are similar enough to express the semantic information and their labels are 1. The training process can be written as
\begin{equation}
	\begin{aligned}
	&(\hat{\bm{\alpha}}_{{\rm sim},1},\hat{\bm{\alpha}}_{{\rm sim},2},\hat{\bm{\theta}}_{{\rm sim},1},\hat{\mathbf{W}}_{\rm sim}) \\&=\mathop{\arg\min}\limits_{{\bm{\alpha}}_{{\rm sim},1},{\bm{\alpha}}_{{\rm sim},2},{\bm{\theta}}_{{\rm sim},1},{\mathbf{W}_{\rm sim}}}L_{\rm CE}\left(label(\mathbf{s},\hat{\mathbf{s}}), Sim32(\hat{\mathbf{s}},\hat{\mathbf{b}}_{sim})\right).
\end{aligned}
\end{equation}

After training, 32 bit CRC can be replaced with Sim32. Similar to CRC, Sim32 needs to judge the similarity from the estimated sentence and only 32 encoded bits  from the true sentence. The transmission is considered successful when $Sim32(\hat{\mathbf{s}},\hat{\mathbf{b}}_{\rm sim})>0.5$. 

\section{Numerical Results}
	In this section,  we will present the numerical results of different  frameworks  and   discuss  the pros and cons of the semantic-based HARQ. We will also compare their bit consumption  with competing ones.

\subsection{Configurations of the simulation system}
The English version of the proceedings of the European Parliament\cite{koehn2005europarl} is chosen as the dataset, which has 2.2 million sentences and 53 million words. We set up the dictionary using 30,000 most common words. Thus, these words can be denoted by 2-byte integers.  The length of the sentences is restricted  between 4 and 30. After this pre-processing, these sentences are split into a training set of 500,000 sentences and a test set of 50,000  sentences. The input sentences $\mathbf{s}$ are padded to their max length $L=30$ with zeros, and the number of units in the hidden layers $M$ is 128. The detailed settings of the proposed networks are shown in Table \ref{Tarh}. 

 	\begin{table}[!h]
	\centering	
	\caption{Settings of the proposed Networks. }
	\begin{threeparttable}

		\begin{tabular}{c|c|c|c|c}    %
			\toprule
			Networks	& Modules & Layer & Output & Activation\\
			& & & dimensions & function  \\ \hline
			
			\multirow{14 }{*}{SC}& INPUT &$\mathbf{s}$  &	 30 & / \\ \cline{2-5}
			&\multirow{3 }{*}{$SC_{\rm en}$}  & Embedding  &	 (30,128) & None \\
			&	& 6 $\times$ Transformer &	 (30,128)& None \\
			& & FC &	 (30,32)& ReLU \\ \cline{2-5}
			
			&\multirow{3 }{*}{$Q$} 
			&    FC & 	(30,30) &ReLU \\ 
			&&   1-bit quantization & 	(30,30) &/ \\ 
			&&   Mask & 	($L_{s}$,30) &/ \\ 
			\cline{2-5}

			&\multirow{2 }{*}{$Q^{-1}$} 
			&    Padding & 	(30,30) &\ \\ 
			&&   FC & 	(30,32) &ReLU \\ \cline{2-5}

			&\multirow{3}{*}{$SC_{\rm de}$}  &FC &	 (30,128)& ReLU \\
			& &6 $\times$ Transformer &	 (30,128)& None \\
			& & FC &	 (30,30000)& SoftMax \\ 
			\cline{2-5}
			& OUTPUT&$\hat{\mathbf{s}}$   &	 30& Argmax \\

			\bottomrule
			\toprule

			\multirow{10 }{*}{SCHARQ}& INPUT &$\mathbf{s}$&	30  & / \\ \cline{2-5}
			&\multirow{4 }{*}{Encoder $i$} & $SC_{\rm en}$ &	(30,32) &   \\
			
			& & Reshape &	 960& / \\
			& & FC &	 $b_i$& ReLU \\
			&&   1-bit quantization & 	 $b_i$ &/ \\ \cline{2-5}
			&	\multicolumn{4}{c}{bit error channel} \\ \cline{2-5}

			&\multirow{3 }{*}{Decoder $i$} & FC &	 960& ReLU \\
			& & Reshape 
			&(30,32)& ReLU \\
			& & $SC_{\rm de}$ &	(30,30000) &SoftMax \\

			\cline{2-5}
			& OUTPUT&$\hat{\mathbf{s}}_i$   &	 30& Argmax \\ 
			\bottomrule

			\bottomrule
		\end{tabular}
		\begin{tablenotes}
			\item{*} For convenience, $SC_{\rm en}$ is directly used in the layer column to represent the same layers  shown in the $SC_{\rm en}$ module of the SC. Sim32 only consists of the proposed modules and some FC layers; thus, its detailed architecture is omitted here.
		\end{tablenotes}
	\end{threeparttable}
	\label{Tarh}
\end{table}

 Huffman coding is used as the conventional source coding, and the average length of coded bits are compared with the SC  in Table \ref{T1}. Due to different lengths of sentences, the SC is still not efficient compared with the Huffman code. Meanwhile, the SC has no  mechanism to guarantee no error, and it costs numerous  bits to reduce WER further when WER is already very small. For example,  SC  with $B=40$ needs 1/3 more bits than that with  $B=30$, but  WER only decreases by nearly 1\%. In the following simulations, $B=30$ is chosen  to balance bit consumption and WER. The semantic encoder with fixed bit length, called fixed SC,  encodes the sentence into 500 bits. The WER of the fixed SC is much higher than the SC with an average of 490 bits per sentence because it is weak in dealing with long sentences.

 	\begin{table}[!h]
	\centering	
	\caption{SC source coding with different $B$. }
	\begin{tabular}{>{\sf }c|c|c|c}    %
		\toprule
		& $B$ & Bits/Sentence & WER\\ \hline
		\multirow{2}{*}{SC}	&40&	 654 & 0.01\% \\
		
		& 30 &	 490 & 1.03\% \\ \hline
		Fixed SC &/& 500&9.81\%\\ \hline
		Huffman	&/&397	 & / \\	
		\bottomrule
	\end{tabular}
	\label{T1}
\end{table}

Binary symmetric channel,  where  transmit bits are randomly inversed, is used. The WER and SER are analyzed along with the change in BER. 
 The transmission using Huffman source coding, RS channel coding, and HARQ is  called Huffman-RS-HARQ, where the number of the SC coders is $R=4$, and  the code rates of each transmission are $\frac{k}{n_1}=\frac{5}{7}$, $\frac{k}{n_2}=\frac{5}{11}$, $\frac{k}{n_3}=\frac{5}{15}$ and $\frac{k}{n_4}=\frac{5}{19}$. Thus,  $n_R=1508$.  The series and parallel SC-RS-HARQ methods only transmit three times at most and the corresponding $n_R=490\times \frac{n_3}{k}=1470 $, which is lightly fewer than the conventional one.  For  SCHARQ,  $R=3$ and the  code lengths  are set as $n_1=300$, $n_2=500$, and $n_3=1000$, respectively.

\subsection{Performance of SC Combined with Conventional Methods}

	\begin{figure}[!h]
		\centering
		
			\subfloat[ ]{
			\includegraphics[width=4in]{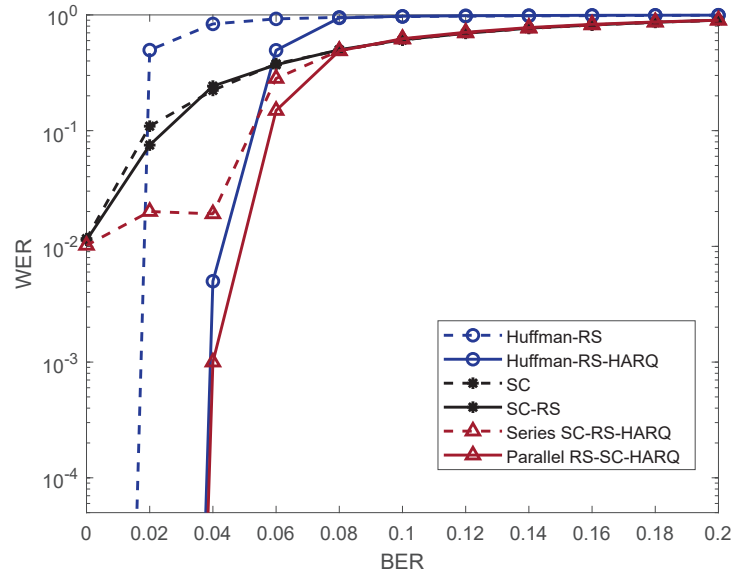}}\\
		\subfloat[ ]{
			\includegraphics[width=4in]{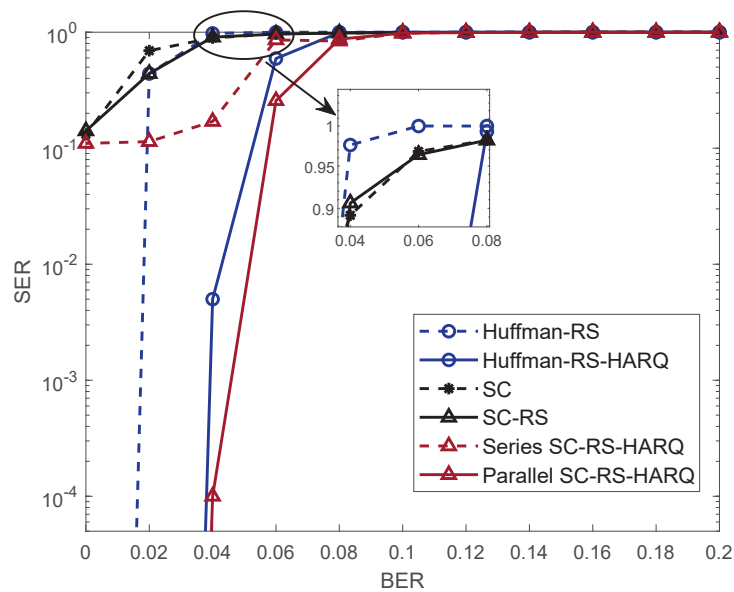}}
		
		\caption{(a) WER performance of the different semantic  competing conventional methods. (b) SER performance of the different semantic  competing conventional methods. }
		\label{SC}
	\end{figure}

In Fig. \ref{SC}(a), Huffman-RS represents  the first transmission of Huffman-RS-HARQ with code rate  $\frac{5}{7}$ and code length  of approximately 555 bits.  The  Huffman-RS has the worst capability to cope with the change in BER.   SC requires fewer bits (approximately 490 bits) but has better performance than the Huffman-RS, especially when BER is high. This phenomenon demonstrates the SC  can handle  bit errors even if it does not learn to do it. The series and parallel SC-RS-HARQ both perform better than  Huffman-RS-HARQ   when BER $>$ 0.04. However,  the Huffman-RS-HARQ and parallel SC-RS-HARQ can guarantee almost zero WER when BER $<$ 0.04 because the estimated sentences are directly decoded by the conventional method, whereas   the series SC-RS-HARQ always has tiny errors when testing. Due to the SC source coding, the series and parallel SC-RS-HARQ is better than the Huffman-RS-HARQ when BER $\geq$ 0.06, and its performance is the same as the SC  when BER $\geq$ 0.08. Therefore, the semantic-based SC is still effective when the conventional method cannot work.

In Fig. \ref{SC}(b), the SC has a higher SER than the Huffman-RS when BER $<$ 0.03 because  nearly 1\%  of the error words randomly spread out   the estimated sentences and causes about 10\% of estimated sentences with one or two error words even  when the conventional Huffman-RS-HARQ has no errors.  The series SC-RS-HARQ also has the same phenomenon, and its SER can only reach 0.1 while its WER is around 0.01. In contrast, the conventional methods  can estimate sentences perfectly if the number of  wrong bits is below  their error correction capability.  The parallel SC-RS-HARQ can correct the error words after the SC decoder; thus, it can achieve perfect transmission when BER$<$ 0.04.  Meanwhile, the SER of the parallel SC-RS-HARQ always decreases earlier with   BER than the competing methods, which shows its superiority when  facing high BER conditions.

\begin{figure}[!h]
	\centering

		\includegraphics[width=4in]{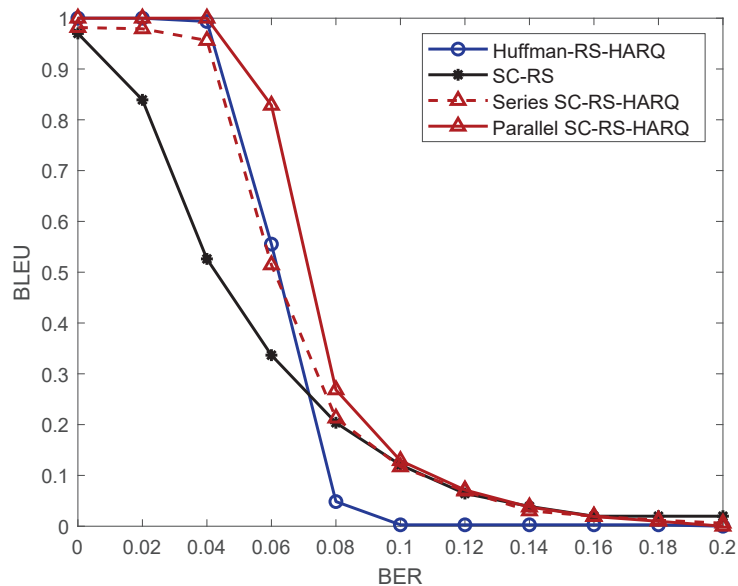}

	\caption{BLEU performance of the proposed SC-RS-HARQ methods and conventional Huffman-RS-HARQ method.  }
	\label{SCBLEU}
\end{figure}

We also compare the BLEU performance of these methods in Fig. \ref{SCBLEU}.  Similar to their SER performance, the parallel SC-RS-HARQ always has the best BLEU performance while that of the series SC-RS-HARQ is worse than that of the conventional Huffman-RS-HARQ when BER $\leq$  0.06. However, the performance gap between the series SC-RS-HARQ and the  Huffman-RS-HARQ is smaller than that in Fig. \ref{SC}(b) because the wrong sentences still reserve some correct semantic information. The three SC-based methods have the same performance when BER $>$ 0.1 because the SC network plays a major role in high BER. Overall, the series SC-RS-HARQ is not too bad under the  BLEU measurement because the reserved tiny error words  when BER $<$ 0.03 have little impact on its semantic information.

In the above simulation, the semantic methods  have better WER performance than the conventional methods  if they are combined with conventional RS coding and HARQ properly. The SER performance of parallel SC-RS-HARQ is guaranteed to surpass the conventional Huffman-RS-HARQ while that of the series one  has worse SER. Overall, the proposed methods are easily applied in the conventional HARQ systems, and the performance  is improved for high BER.

\subsection{Performance of SCHARQ}

	\begin{figure}[!h]
	\centering
	
	\subfloat[ ]{ 
		\includegraphics[width=3.5in]{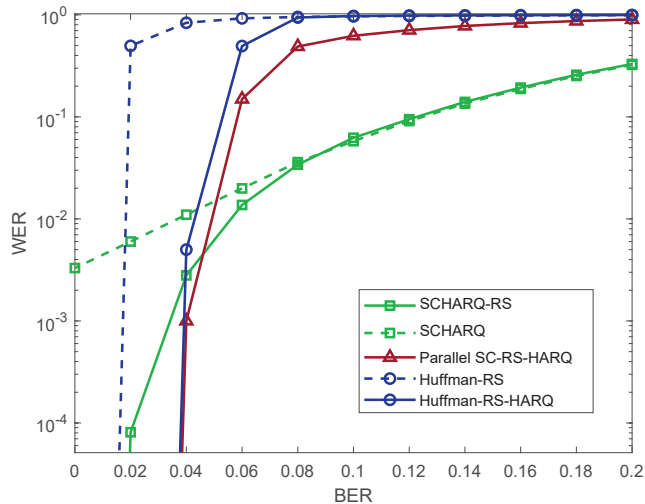}}\\
	\subfloat[ ]{
		\includegraphics[width=3.5in]{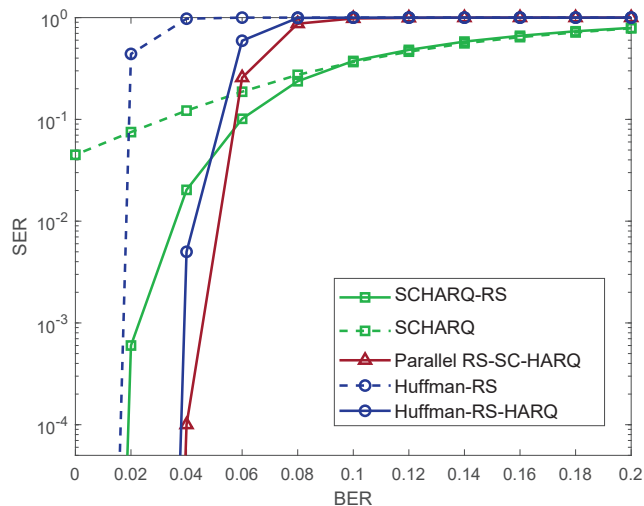}}
	
	\caption{(a) WER of the proposed SCHARQ  and the competing methods. (b) SER of the proposed SCHARQ and the competing methods. }
	\label{SCHARQ}
\end{figure}
	
 In Fig. \ref{SCHARQ}(a), SCHARQ  methods show their superiority when  BER is between 0.04 and 0.2, and increasing the transmit bit limit is helpful to improve the WER of SCHARQ. Because avoiding the tiny error is difficult   when BER $\leq$ 0.04, we also transmit extra parity bits coded by the RS encoder to correct the estimated sentences similar to the parallel SC-RS-HARQ. The SCHARQ with the RS code is called SCHARQ-RS, which has $\frac{5}{7}$ code rate for RS coding.   Similarly, the SCHARQ methods  always have better capability to cope with high BER, as shown in Fig. \ref{SCHARQ}(b). $n_R=1000$ is adequate for SCHARQ to approach almost zero WER when BER $\leq$ 0.04, but its SER can only reach 0.05 at most due to tiny WER.  The conventional RS code helps improve the performance and guarantees the perfect transmission when the error in the SCHARQ does not surpass the capability of the RS code. Thus, SCHARQ-RS can reach 0 WER and 0 SER when BER=0 and surpass SCHARQ when BER $\leq$ 0.1. However, the error that appears in the redundancy bits of the RS code may also misapprehend the correct estimated sentences, which makes SCHARQ-RS become a little worse than SCHARQ when BER is between 0.1 and 0.2.  The parallel SC-RS-HARQ is the best method to use when BER $\leq$ 0.04, where the SER is  guaranteed to be zero, but it performs worse than the SCHARQ methods when BER $>$ 0.06.


\begin{figure}[!h]
	\centering

		\includegraphics[width=4in]{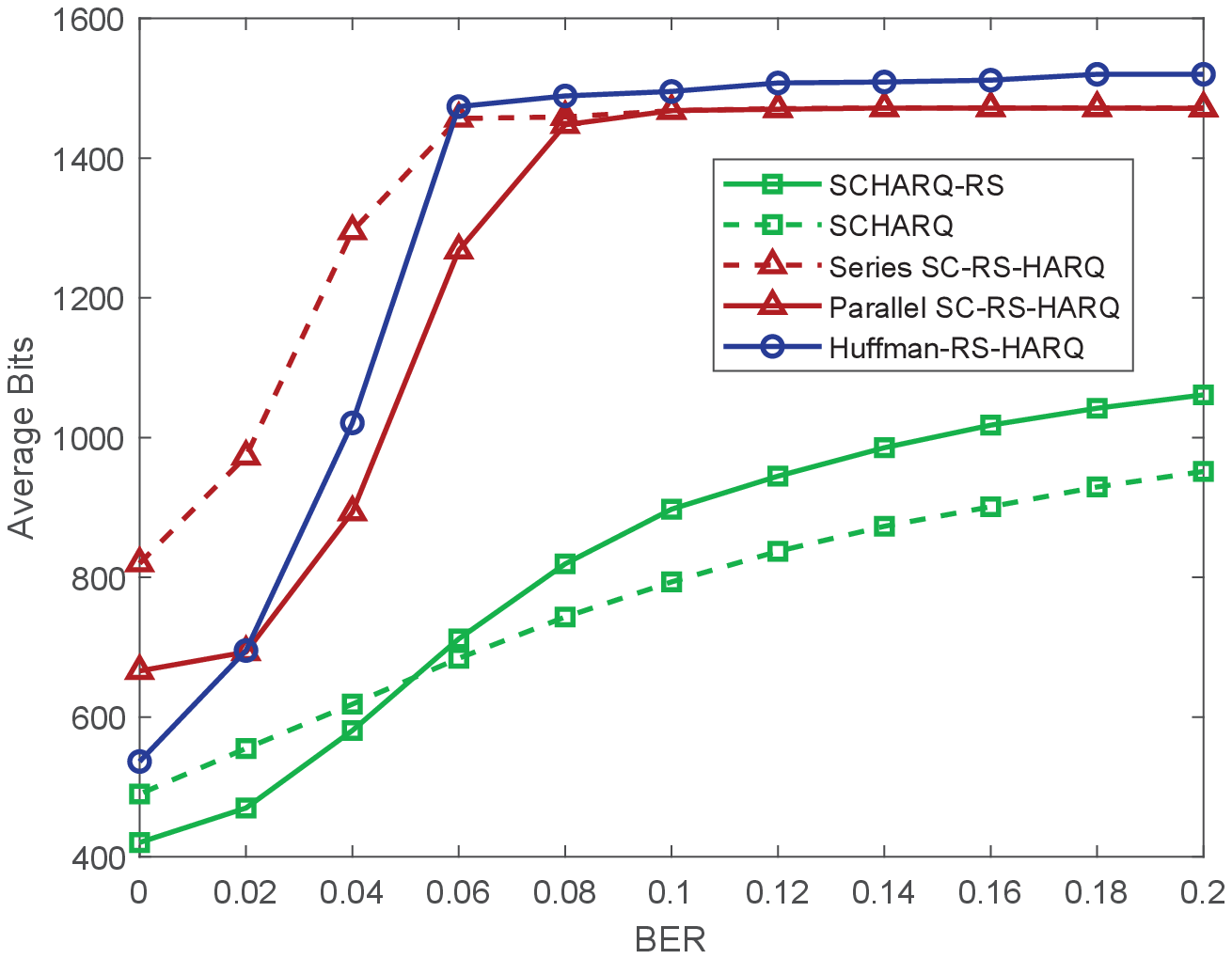}
	
	\caption{Average bit consumption per sentence of different methods. }
	\label{SCHARQ_BC}
\end{figure}

In Fig. \ref{SCHARQ_BC}, the average required number of bits for a sentence transmission are compared.  The conventional Huffman-RS-HARQ and series SC-RS-HARQ reach the  upper limit of transmission times at BER=0.06, and thus its SER is close to 1 when BER$ \geq$0.08. In contrast, the parallel SC-RS-HARQ performs better when BER is low and reaches bit consumption limit at BER=0.08. The series SC-RS-HARQ requires more bits than the parallel one when BER $\leq$ 0.06 because  the series one cannot reach zero SER, and some error sentences need to be retransmitted. The  SCHARQ methods show the superiority of the joint design of semantic source-channel coding and HARQ in reducing transmit bit consumption, especially when BER is high. Especially, SCHARQ-RS needs fewer bits than SCHARQ when BER $\leq$ 0.05  even though it transmits extra redundancy bits. This phenomenon is owed to the reduced times of retransmission with the help of RS code when BER is low. However, the redundancy bits of the RS code cannot bring benefit for high BER  but introduces more errors (Fig. \ref{SCHARQ}(b)); thus, the SCHARQ-RS needs more bits than the SCHARQ when BER=0.05 - 0.2.

From the above discussion, we find the joint design of source-channel coding and HARQ has a significant advantage in reducing bit consumption and improving the SER performance under high BER. The proposed methods show their superiority at different BER scales. However, these designs cannot make full use of the semantic coder because the retransmission decision is still CRC that needs the estimated sentences to  error-free.   Although the semantic method cannot surpass the conventional methods if the BER is low and no error transmission is needed, it brings the possibility  to protect the sentence meaning when some words are  incorrect. In the following, CRC  is replaced with similarity detection to study the pros and cons of  semantic transmission.


\subsection{Pros and Cons of Similarity Detection}
\begin{figure}[!h]
	\centering

		\includegraphics[width=5in]{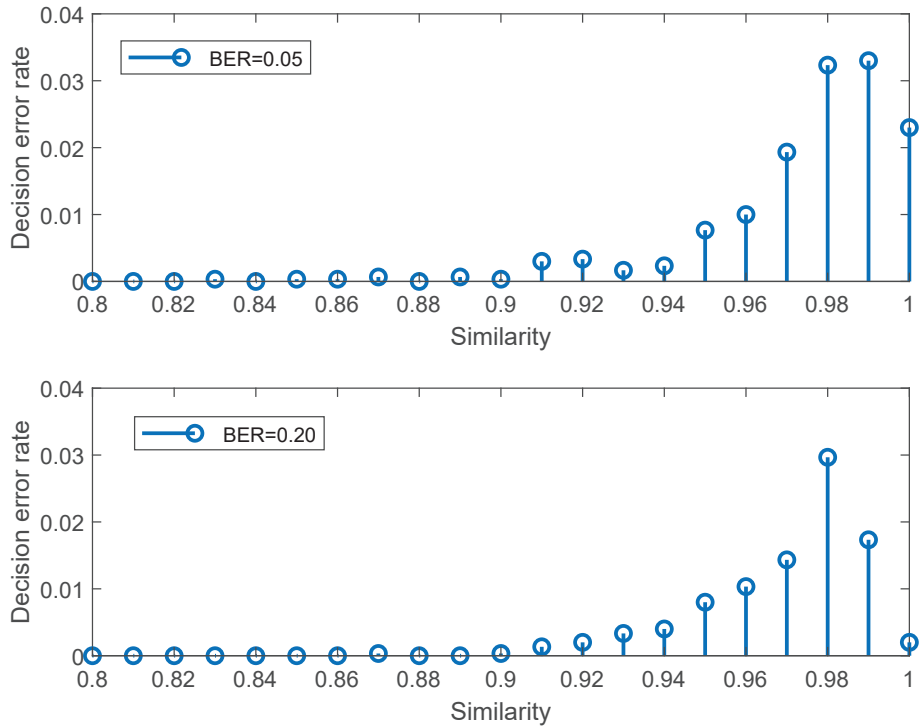}

	\caption{Decision error rate of the Sim32 error detection method. }
	\label{simdec}
\end{figure}

We first show the  decision error rate of Sim32  in Fig. \ref{simdec}. We calculate the similarity according to  (\ref{simcal}) of  50,000 estimated sentences  under two different BER settings. The decision error means the estimated sentences with similarity  larger than 0.98 have Sim32 decision of 0, and those with similarity less than 0.98  have Sim32 decision of 1. As shown in Fig. \ref{simdec}, the error rate is high in the adjacent area of 0.98, which demonstrates that  Sim32 cannot obtain an accurate similarity  because the semantic information of the true sentence is compassed into 32 bits in the transmission. In general, this similarity detection efficiently refuses the estimated sentences with a similarity less than 0.9 and is robust to the change in BER. However, approximately 2\% of the correct sentences are mistaken as dissimilar sentences by Sim32 under BER=0.05. To solve this issue, we only use Sim32 to find similar sentences after CRC detection, and it ensures that the correct sentences  are directly received at the CRC process, and is called CRC-Sim32.

\begin{figure}[!h]
	\centering

		\includegraphics[width=4in]{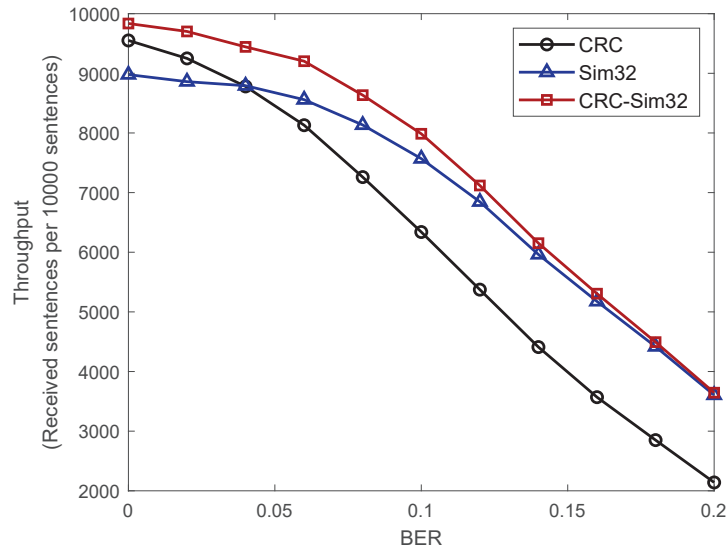}

	\caption{Received sentences from 10000 transmit sentences with different error detection methods.  Only the error detection method at the last transmission  is replaced.  }
	\label{sim_SR}
\end{figure}

In Fig. \ref{sim_SR}, the first two transmissions of SCHARQ still use CRC error detection,  and that of the last transmission is changed into  Sim32, and CRC-Sim32. Thus, the change in  error detection has no influence on reducing retransmission times. The CRC-Sim32 uses CRC  first and then uses Sim32 to decide the sentences that fail in CRC. The CRC-Sim32 needs extra 32 bits but can ensure that the correct sentences are not considered  dissimilar sentences.   The detected error sentences means that these estimated sentences cannot pass the decision after the last transmission, and this error rate is shown in Fig. \ref{sim_SR}. The performance of CRC can represent the SER performance of SCHARQ in Fig. \ref{SCHARQ}(b). The Sim32 detection at the last transmission allows that some estimated sentences with error words are received as similar sentences. This method  increases the number of the received sentences, especially when BER is high. However, some correct sentences are mistakenly decided as dissimilar sentences. Thus, Sim32 receives fewer  sentences than CRC when BER $\leq$ 0.04.  CRC-Sim32 performs best when BER is low and approaches  Sim32 when BER=0.2  because only few correct sentences are mistakenly rejected by Sim32 when BER=0.2.

 Replacing CRC with CRC-Sim32 at all transmission can reduce the bit consumption further by reducing the times of retransmission. For example, the received sentences can be increased by 18\%, and the average bit consumption is reduced by nearly 40 bits when replacing CRC with CRC-Sim32 at all transmission when BER=0.2.

\begin{figure}[!h]
	\centering

		\includegraphics[width=4in]{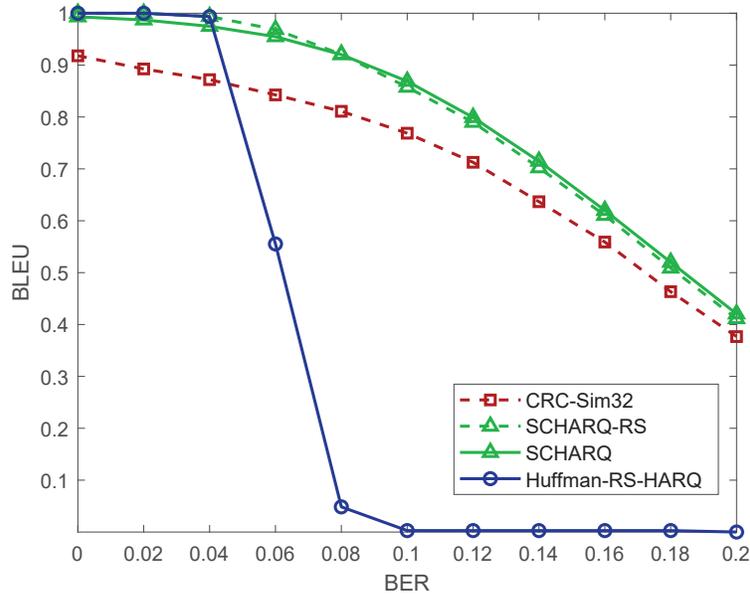}

	\caption{BLEU performance of the proposed joint source-channel coding and HARQ methods. CRC-Sim32 means that BLEU performance of SCHARQ, whose the error detection methods of all transmissions  are replaced with CRC-Sim32.}
	\label{simBLEU}
\end{figure}

Finally, the BLEU performances of SCHARQ-based methods are compared in Fig. \ref{simBLEU}. The SCHARQ-RS, SCHARQ, and Huffman-RS-HARQ use CRC error detection while CRC-Sim32 is the SCHARQ that uses CRC-Sim32 for all the transmission times. Although SCHARQ-RS has an obvious SER performance gap with SCHARQ in Fig. \ref{SCHARQ}(b), they have similar BLEU performance because the few error words that can be corrected by RS code have little influence on the semantic information. Thus, CRC-Sim32 only needs the received sentences to be understood rather than a high BLEU score, thereby causing  some  sentences to need fewer bits than the SCHARQ with CRC error detection. The performance gap between SCHARQ with CRC and  CRC-Sim32  under low BER is larger than that under high BER because  more received sentences have a chance to reach a higher BLEU score under low BER.

The proposed similarity detection aims at adopting an estimated sentence with error words  as long as its semantic meaning is unchanged. However,  similarity detection is such a difficult task because the true sentences are unknown to the receiver. A lot of work must still be done  before the similarity detection becomes reliable in the sentence transmission. 
\subsection{Similar Sentences Received by  Similarity Detection}
 	\begin{table}[!h]
	\centering	
	\small
	\caption{The sentences pass the Sim32  but contain error words. }
	\begin{tabular}{>{\sf }c|p{13cm}}    %
		\toprule
		\multicolumn{2}{c}{BER=0 (CRC)}\\ \hline
		\multirow{2}{*}{1}& RX: this \textbf{aviation} means a lot to eastern and central european countries also for historical reasons because it provides the opportunity to travel freely without barriers \\ 
		&TX: this \textbf{acquis} means a lot to eastern and central european countries also for historical reasons because it provides the opportunity to travel freely without barriers \\ \hline
		\multirow{2}{*}{2}& RX: today we must welcome an important result which is that we consider the citizens of bulgaria and romania to be among those who can benefit from this fundamental \textbf{that}\\
		&TX: today we must welcome an important result which is that we consider the citizens of bulgaria and romania to be among those who can benefit from this fundamental \textbf{instrument}\\
		 \hline
		 
		\multirow{2}{*}{3}& RX: the problem must be treated at its roots by providing better employment and education opportunities for them this internal migration pressure will also \textbf{reconciliation} within the european union\\
		&TX: the problem must be treated at its roots by providing better employment and education opportunities for them this internal migration pressure will also \textbf{decrease} within the european union\\
		\hline
			\multicolumn{2}{c}{BER=0.2 (CRC)}\\ \hline
		\multirow{2}{*}{1}& RX: the \textbf{disputes} just do not back this idea up either\\
		&TX: the \textbf{facts} just do not back this idea up either\\
		\hline
			\multirow{2}{*}{2}& RX: romania has invested more vote eur 1 billion and the results have undertaken countries positive in all \textbf{this communities are}\\
			&TX: romania has invested more than eur 1 billion and the results have definitely been positive in all \textbf{the evaluation reports}
			\\ \hline
			
			\multirow{2}{*}{3}& RX: no one \textbf{are} interested \textbf{freely} in a two track europe\\
		&	TX: no one \textbf{is} interested \textbf{nowadays} in a two track Europe\\ \hline
		
			\multicolumn{2}{c}{BER=0 (CRC-Sim32)}\\ \hline
			
		\multirow{2}{*}{1}& RX:	it was already established in 2007 that once the technical \textbf{negotiations} were fulfilled bulgaria and romania would join the schengen area in \textbf{2008}\\
		&	TX:	it was already established in 2007 that once the technical \textbf{criteria} were fulfilled bulgaria and romania would join the schengen area in \textbf{2011}\\ \hline
		
	\multirow{2}{*}{2}& RX:	fellow members bulgaria and romania have completed their \textbf{development} and we are building on \textbf{greater} security systems in cooperation with \textbf{our} schengen partners\\
	&	TX:	fellow members bulgaria and romania have completed their \textbf{job} and they are building on \textbf{these} security systems in cooperation with \textbf{their} schengen partners\\ \hline
	
	\multirow{2}{*}{3}& RX: what is the situation when it comes to the judicial reform and anti corruption measures that are still \textbf{achieved}\\
&TX:	what is the situation when it comes to the judicial reform and anti corruption measures that are still \textbf{needed}\\ \hline

	\multicolumn{2}{c}{BER=0.2(CRC-Sim32)}\\ \hline
\multirow{2}{*}{1}& RX: bulgaria and romania are absolutely not \textbf{preparations} for schengen\\
&TX: bulgaria and romania are absolutely not \textbf{ready} for Schengen\\ \hline
\multirow{2}{*}{2}& RX: we \textbf{are on} double standards\\
&TX: we \textbf{cannot allow} double standards\\ \hline

\multirow{2}{*}{3}& RX: what \textbf{president} is the commission giving that this \textbf{rapporteur} will be \textbf{resolved} effectively\\
&TX: what \textbf{guarantees} is the commission giving that this \textbf{problem} will be \textbf{tackled} effectively\\
\bottomrule
\end{tabular}
\label{T2}
\end{table}

Here, we will analyze some common mistakes that appear in the "similar sentences" decided by the proposed  methods. As shown in Table \ref{T2}, some similar sentences of SCHARQ are collected under four settings. Here, CRC in the bracket  means the first two transmissions are decided by CRC, and the last transmission is detected by CRC-Sim32. CRC-Sim32 means that all the transmissions are  detected by CRC-Sim32. TX is the transmitted sentence, and RX is the received sentence after SCHARQ. Three pairs of TX and RX sentences are shown under different settings.

Most of the similar sentences under BER=0 (CRC)  contain only one or two error words. Meanwhile, these mistakes only happen in long sentences  because the 1000-bit limitation of the semantic network may not be enough for these sentences. These three received sentences demonstrate that the wrong nouns are difficult to judge for similarity detection because replacing a noun usually has no influence on the grammar. For Sentence 1, ``acquis" is replaced with ``aviation", which may associate with ``travel freely without barrier". For Sentence 2, the noun ``instrument" becomes pronoun ``that" and its meaning is vague. For Sentence 3, its meaning has no change but its grammar contains a mistake. Pressure ``reconciliation"  can also be considered  pressure ``decrease".

When BER=0.2 (CRC), more wrong words appear in similar sentences, and the similarity detection may also make  mistakes. For example, Sentence 2 is decided as a similar sentence, but its meaning has been damaged by the error words.  Sentence 1  also has wrong noun and  its meaning is changed to some extent. Meanwhile,  Sentence 3  can be understood.

The use of CRC-Sim32 as a retransmission decision saves bit consumption further but introduces more similar sentences with more wrong words than  CRC. When BER=0, Sentences 1 and 2 face the change of words, and these words may lead a serious misunderstanding. For example, the year ``2011" is replaced with ``2008" in Sentence 1, and this mistake directly affects the behavior of the receiver. This phenomenon  also  exists when BER=0.2. Thus, similarity detection is still not very reliable. Although  similar sentences with error words replaced with synonyms are received, some sentences with changing semantic information are also  difficult to be found and rejected.

In general, the proposed Sim32 is a good attempt to exploit the capability of semantic architectures, which can protect the semantic information when facing mistakes and trying to repair the sentences according to semantic relation. However, some similar sentences decided by Sim32 may still lead to misunderstanding   
  
\section{Conclusions}
In this paper, we have investigated  semantic coding. We have combined it  with conventional RS channel coding and IR-HARQ and developed two different frameworks, namely, series SC-RS-HARQ and parallel SC-RS-HARQ. By comparing the two proposed frameworks  and conventional methods, we find that the semantic encoder has better performance when facing high BER. However, it cannot guarantee  error-free transmission. The parallel SC-RS-HARQ exploits the different advantages of  the semantic architecture and the conventional method and outperforms the conventional IR-HARQ method.
We have also designed a joint source-channel coding and HARQ framework called SCHARQ. This framework is more flexible and efficient because  it can transmit incremental bits to solve the issues of different sentence lengths and varying channel conditions. Thus, it has the best performance among the other competing methods when BER is high but a little weaker when BER is low.
To exploit the full potential of the semantic coder, we have   proposed a similarity detection called Sim32 to detect the semantic error in the estimated sentences  and combined it with CRC  called CRC-Sim32. The proposed error detection methods  allow similar sentences to be received so that  more sentences can be transmitted, especially when BER is high. However, some sentences with changing semantic information are still mistakenly received. In the future, more work is needed to improve its reliability.

	\bibliographystyle{IEEEtran}
	\bibliography{bibtex0320}
	
	%
	
	
	
	

\end{document}